\documentstyle [12pt]{article}

\def\b{\begin{equation}} \def\e{\end{equation}}
\def\bd{\begin{displaystyle}} \def\ed{\end{displaystyle}}
\def\ba{\begin{array}} \def\ea{\end{array}}

\def\bee{\begin{enumerate}}
\def\eee{\end{enumerate}}

\def\bes{\begin{eqnarray*}}
\def\ees{\end{eqnarray*}}
\def\be{\begin{eqnarray}}
\def\ee{\end{eqnarray}}

\begin{document}
\title{ Euler-Heisenberg lagrangian through Krein regularization}

\author{A. Refaei$^{}$ \thanks{e-mail:
abr412@gmail.com}}

\maketitle \centerline{\it Department of Physics,  Sanandaj branch, Islamic Azad
University, Sanandaj, Iran.}

\begin{abstract}
The Euler-Heisenberg effective action at the one-loop for a constant electromagnetic field  is derived in Krein space quantization with Ford's idea of fluctuated light-cone. In this work we present a perturbative, but convergent solution of the effective action. Without using any renormalization procedure, the result coincides with the famous renormalized Euler-Heisenberg action.

\end{abstract}

\section {Introduction}
A well known problem of quantum field theory is the presence of the ultraviolet
divergences that appear by practical calculations in perturbation theory.
 Because of the singular behavior of the Green functions at
small distances they are exhibited by the divergence of many integrals (over coordinates
and momentums). A consistent scheme for eliminating the ultraviolet divergences and obtaining
finite results is the theory of renormalizations \cite{itzu,wehais}, that can be
carried out consequently in renormalizable field theories. According to this rule the counterterms that have the structure of the individual terms
of the classical action are intoduced. They are interpreted in terms of renormalizations of
the fields, the masses and the coupling constants.

Some new parameters such as a dimensionless
regularizing parameter $\alpha$ and a dimensionful renormalization parameter
$\mu$ could be introduced. After subtracting the divergences and going to the limit $\alpha \rightarrow 0$ the regularizing
parameter disappears, but the renormalization parameter $\mu$ remains
and enters the finite renormalized expressions. This process in the eliminating of divergences is known as regularization. In renormalized quantum
field theories the change of this parameter is compensated by the change of
the coupling constants of the renormalized action, that are defined at
the renormalization point characterized by the energy scale $\mu$. The physical
quantities are renormalization invariant \cite{iga}.

 Our method is different from the usual one. We introduce a way to gain a result that doesn't have any divergences and the theory is renormalized automatically \cite{gazeau00,ta3,ta4,rota,gagarota,gaj,sb,derotata,taj,reta}. To obtain such theory and in order to remove the singularities, we should modify the Green function \cite{rota,reta,reta2}. The basic idea of our method is as follow:
\begin{quote}
 The singular behavior of Green function at short relative distances (ultraviolet divergence)
or in the large relative distances (infrared divergence) leads to divergences of the QFT. The
ultraviolet divergence appears in the following terms of Green function in the limit $\sigma=0$:
$$\frac{1}{\sigma}, \quad \ln\sigma \quad and \quad \delta(\sigma).$$
It was conjectured long ago \cite{des,dew} that quantum metric fluctuations might smear out the
singularities of Green functions on the light-cone {\it{i.e.}} $\delta(\sigma)$. Along with this line the model
described by Ford \cite{ford97}, which is based on the quantum metric fluctuations, does smear out the
light cone singularities, but it does not remove other ultraviolet divergences of quantum field
theory.

 The quantum field theory in Krein space, {\it{i.e.}} indefinite metric quantization, studied previously
for other problems \cite{dir,rami}, was utilized for the covariant quantization of the minimally
coupled scalar field in de Sitter space \cite{gazeau00}. In this method, the auxiliary negative norm states
(negative frequency solution) have been utilized, the modes of which do not interact with the
physical states or real physical world. One of the interesting results of this construction is
that the quantum field theory in Krein space removes all ultraviolet divergences of quantum
field theory with exception of the light-cone singularity \cite{gazeau00,ta3,ta4}.
\end{quote}
The essential point of this paper is the unavoidable presence of the negative norm states with quantum metric fluctuatiton, they play a renormalizing role.
Consideration of the negative norm states was proposed by Dirac in 1942 \cite{dir}. In 1950, Gupta
applies the idea in QED \cite{gu}. The presence of higher derivative in the Lagrangian also lead to
ghosts, states with negative norm \cite{haw}. Mathematically, for preserving the covariant principle, the
auxiliary negative norm states were presented. Their presence has also different consequences,
for example in QED the negative energy photon disappears \cite{gu}, and in de Sitter the infrared
divergence eliminated \cite{gazeau00}. The physical interpretation, however, is not yet clear and any further
interpretation needs far more investigations \cite{rami,ha,viba,bl}. In 1985, Allen showed that a covariant quantization of minimally coupled scalar field cannot be constructed from positive norm states alone \cite{allen85}.

In this paper, we employ the derivation rules of the low-energy effective action for QED. In
gauge theories, the one-loop effective action has been calculated exactly and analytically only for very special
gauge field backgrounds, such as those with constant field strength,
based on the seminal work of Heisenberg and Euler \cite{heeu}, and Schwinger \cite{j}; and computing numerically in the small mass limit \cite{geduad} also for some very special one-dimensional
cases where the field is inhomogeneous \cite{du}. Our purpose is to extend the Krein space method and quantum metric
fluctuation at the linear approximation,
for which we can compute the renormalized one-loop effective action. In the Euler
Heisenberg case, divergences occur only in the
coefficients of the terms which are zeroth order and second order in field
strengths, which are cured by the mass and charge renormalizations normally.
But, we have shown that the result of our method not only
doesn't have any singularities but also is equivalent to the renormalized result which has been reported by Euler and Heisenberg  and by Schwinger long time ago.

The present work is organized as follows: In section 2, we briefly recall the propagator derivation for scalar field in Krein space
quantization with Ford's idea of metric fluctuation. Section 3 is devoted to express the one-loop
effective action in terms of the new propagator in our method which is expressed by the Principal Part.
Then, we utilize the Schwinger's method and other techniques for
generating the perturbative expression in section 4, we also present
the main results of our letter. In appendices, we provide some details of useful and necessary calculation techniques.

\setcounter{equation}{0}
\section{Feynman propagator in Krein space with fluctuation of the light-cone}

 The primary concepts of  quantization in Krein space are introduced: The scalar Green function is constructed and by utilizing the quantum fluctuation of light-cone, the Feynman propagator in the new approach will be reproduced.

 A classical scalar field $\phi(x)$ satisfies the
following field equation  \b \label{eq:1}
(\Box+m^2)\phi(x)=0=(\eta^{\mu\nu}\partial_\mu
\partial_\nu+m^2)\phi(x),\;\;
\eta^{\mu\nu}=\mbox{diag}(1,-1,-1,-1).\e
 Inner ({\it Klein-Gordon}) product and related norms are defined by \cite{bida}
\b(\phi_1,\phi_2)=-i\int_{t=\mbox{const.}}\phi_1(x)\stackrel{\leftrightarrow}
{\partial}_t\phi_2^*(x)d^3x.\e Two sets of solutions are given by:
\b u_p(k,x)=\frac{e^{i\vec k.\vec x-iwt}}{\sqrt{(2\pi)^32w}}
=\frac{e^{-ik.x}}{\sqrt{(2\pi)^32w}},\e
\b u_n(k,x)=\frac{e^{-i\vec
k.\vec x+iwt}}{\sqrt{(2\pi)^32w}}
=\frac{e^{ik.x}}{\sqrt{(2\pi)^32w}},\e where $ w(\vec k)=k^0=(\vec
k .\vec k+m^2)^{\frac{1}{2}} \geq 0$, note that $u_n$ has the
negative norm. In Krein space the quantum field is defined as
follows \cite{ta4}:
 \b \phi(x)=\frac{1}{\sqrt
2}[\phi_p(x)+\phi_n(x)],\e
where $$ \phi_p(x)=\int d^3\vec k [a(\vec k)u_p(k,x)+a^{\dag}(\vec
k)u_p^*(k,x)],$$ $$ \phi_n(x)=\int d^3\vec k [b(\vec
k)u_n(k,x)+b^{\dag}(\vec k)u_n^*(k,x)].$$  $a(\vec k)$ and $b(\vec
k)$ are two independent operators. The time-ordered product
propagator for this field operator is \b iG_T(x,x')=<0\mid
T\phi(x)\phi(x') \mid 0>=\theta (t-t'){\cal W}(x,x')+\theta
(t'-t){\cal W}(x',x).\e In this case we obtain \b
G_T(x,x')=\frac{1}{2}[G_F(x,x')+(G_F(x,x'))^*]=\Re G_F(x,x'),\e
where the Feynman Green function is defined by \cite{bida}
\begin{eqnarray}
G_F(x,x')&=&\int \frac{d^4 p}{(2\pi)^4}e^{-ip.(x-x') }\tilde
G_F(p)=\int \frac{d^4p}{(2\pi)^4}\frac{e^{-ip.(x-x')}}{p^2-m^2+i\epsilon} \label{eq:2}
\nonumber\\&=&-\frac{1}{8\pi}\delta
(\sigma_0)+\frac{m^2}{8\pi}\theta(\sigma_0)\frac{J_1
(\sqrt{2m^2\sigma_0})-iN_1 (\sqrt{2m^2\sigma_0})}{\sqrt{2m^2
\sigma_0}}\nonumber\\ & & -\frac{im^2}{4\pi^2}\theta(-\sigma_0)\frac{K_1
(\sqrt{-2m^2\sigma_0})}{\sqrt{-2m^2 \sigma_0}},\end{eqnarray}
 where
$\sigma_0=\frac{1}{2}(x-x')^2 .$ So we have \begin{eqnarray} G_T(x,x')&=&\int
\frac{d^4 p}{(2\pi)^4}e^{-ip.(x-x')}{\cal
PP}\frac{1}{p^2-m^2}\nonumber\\ &=&-\frac{1}{8\pi}\delta
(\sigma_0)+\frac{m^2}{8\pi}\theta(\sigma_0)\frac{J_1
(\sqrt{2m^2\sigma_0})}{\sqrt{2m^2 \sigma_0}}, \;\;x\neq x',
\label{28}
\end{eqnarray}
${\cal PP}$ stands for the principal parts. Contribution of the
coincident point singularity $(x=x')$ merely appears in the
imaginary part of $G_F$ (\cite{ta3} and equation (9.52) in \cite{bida})
$$ G_F(x,x)=-\frac{2i}{(4\pi)^2}\frac{m^2}{d-4}+G_F^{\mbox{finit}}(x,x),$$ where $d$ is the space-time dimension and
$G_F^{\mbox{finit}}(x,x)$ becomes finite as $d\longrightarrow 4$.
Note that the singularity of the Eq.(\ref{28}) takes place only on
the cone \emph{i.e.,} $x\neq x', \sigma_0=0$.

It has been shown that the quantum metric fluctuations remove the
singularities of Green's functions on the light-cone \cite{ford97}.
Therefore, the quantum field theory in Krein space, including the
quantum metric fluctuation $\left(
g_{\mu\nu}=\eta_{\mu\nu}+h_{\mu\nu}\right)$, removes all the
ultraviolet divergencies of the theory \cite{ford97,rota}, so one
can write: \b \langle G_T(x,x')\rangle = -\frac{1 }{8\pi}
\sqrt{\frac{\pi}{2\langle\sigma_1^2\rangle}}
\exp\left(-\frac{\sigma_0^2}{2\langle\sigma_1^2\rangle}\right)+
 \frac{m^2}{8\pi}\theta(\sigma_0)\frac{J_1(\sqrt {2m^2
 \sigma_0})}{\sqrt {2m^2 \sigma_0}},\label{210}\e
where $2\sigma= g_{\mu\nu}(x^{\mu}-
x'^{\mu})(x^{\nu} - x'^{\nu})$ and $\sigma_1$ is the first order shift in $\sigma$, due to
the linear quantum gravity ($\sigma=\sigma_0+\sigma_1+\emph{O}(h^2)$). The average value is taken over the quantum metric fluctuation and in the case of $2\sigma_0 =\eta_{\mu\nu}(x^{\mu}- x'^{\mu})(x^{\nu} - x'^{\nu})=0$ we have
$\langle\sigma_1^2\rangle\neq 0$. So, we get \b \langle
G_T(0)\rangle = -\frac{1 }{8\pi}
\sqrt{\frac{\pi}{2\langle\sigma_1^2\rangle}} +
 \frac{m^2}{8\pi}\frac{1}{2}.\e
It should be noted that $ \langle\sigma_1^2\rangle $ is related to
the density of gravitons \cite{ford97}.

By using the Fourier transformation of Dirac delta function,
$$ -\frac{1}{8\pi}\delta(\sigma_0)= \int \frac{d^4
p}{(2\pi)^4}e^{-ip.(x-x')}{\cal PP}\frac{1}{p^2},$$or equivalently
$$\frac{1}{8\pi^2}\frac{1}{\sigma_0}= -\int \frac{d^4
p}{(2\pi)^4}e^{-ip.(x-x')}\pi\delta(p^2),$$ for the second part of
Green function, we obtain \b
\frac{m^2}{8\pi}\theta(\sigma_0)\frac{J_1(\sqrt {2m^2
 \sigma_0})}{\sqrt {2m^2 \sigma_0}}=\int \frac{d^4
p}{(2\pi)^4}e^{-ip.(x-x')}{\cal PP}\frac{m^2}{p^2(p^2-m^2)}.\e
And for the first part we have $$
-\frac{1}{8\pi}\sqrt{\frac{\pi}{2\langle\sigma_1^2\rangle}}\exp\left[-\frac{(x-x')^4}
{4\langle\sigma_1^2\rangle}\right]=\int \frac{d^4
p}{(2\pi)^4}e^{-ik.(x-x') }  \tilde{G}_1(p),$$
 where $\tilde{G_1}$ is fourier transformation of the first part of the Green function (\ref{210}). Therefore, we
obtain \b \label{eq:22}<\tilde G_T(p)>=\tilde{G}_1(p)+{\cal
PP}\frac{m^2}{p^2(p^2-m^2)}.\e It has been proved
that in the one-loop approximation, the Green function in Krein
space quantization, which appears in the transition amplitude is
\cite{ta4}: \b <\tilde G_T(p)>\mid_{\mbox{one-loop}}\equiv \tilde
G_T(p)\mid_{\mbox{one-loop}}\equiv {\cal PP}
\frac{m^2}{p^2(p^2-m^2)}.
\label{213} \e
 That means in the one-loop
approximation, the contribution of $\tilde{G_1}(p)$ is negligible.
It is worth to mention that in order to improve the UV behavior in
relativistic higher-derivative correction theories, the propagator
(\ref{213}) has been used by some authors \cite{bach,ho,ka}.

\setcounter{equation}{0}
\section{Fermion determinant of one-loop effective action in Krein regularization}

  We start from the general formalism which was presented in \cite{reta2}. The one-loop effective action in QED
reduces to computing the fermion determinant
\begin{eqnarray} J&=& \frac{i}{2} Tr
\ln \left[1-\frac{2eA.p+\frac{e}{2}\sigma_{\mu\nu}
F^{\mu\nu}-e^2A^2}{p^2-m^2+i\epsilon}\right] \nonumber\\
&=&\frac{i}{2}
\int d^4x<x|\ln \left[1-\frac{2eA.p+\frac{e}{2}\sigma_{\mu\nu}
F^{\mu\nu}-e^2A^2}{p^2-m^2+i\epsilon}\right]|x> \label{315}.\end{eqnarray}
 One can write this determinant, in proper time method, as
\cite{itzu}
\begin{eqnarray}
J&=&\frac{i}{2}\int_{0}^{\infty}dss^{-1}e^{-ism^2}Tr\exp\left\{-is[p^2-e(p.A+A.p)-\frac{e}{2}\sigma_{\mu\nu}F^{\mu\nu}+e^2A^2]\right\}\nonumber\\
&-& \frac{i}{2}\int_{0}^{\infty}dss^{-1}e^{-ism^2}Tr \exp(-isp^2)\nonumber\\
&=&\frac{i}{2}\int_{0}^{\infty}dss^{-1}e^{-ism^2}\left[Tr U(s)- Tr
U_0(s)\right], \end{eqnarray}
 where
$U(s)=\exp\left\{-is[p^2-e(p.A+A.p)-\frac{e}{2}\sigma_{\mu\nu}F^{\mu\nu}+e^2A^2]\right\}$
and $U_0(s)=\exp(-isp^2).$ In Krein space quantization including
the quantum metric fluctuation, equation $(3.1)$ reads as \b
J_{kr}=\frac{i}{2} Tr \ln \left[1-(2eA.p+
\frac{e}{2}\sigma_{\mu\nu} F^{\mu\nu}-e^2A^2) {\cal PP}
\frac{m^2}{p^2(p^2-m^2)}\right].\e If we
take$$V=\frac{1}{2}m^2(2eA.p+\frac{e}{2}\sigma_{\mu\nu}
F^{\mu\nu}-e^2A^2),$$ we can write
\begin{eqnarray}J_{kr}&=&\frac{i}{2}Tr\ln
\Big[1- {\cal PP} \frac{2V}{p^2(p^2-m^2)}\Big]\nonumber\\
&=&\frac{i}{2}Tr\ln \Big[1-
V\Big(\frac{1}{p^2(p^2-m^2)+i\epsilon}+\frac{1}{p^2(p^2-m^2)-i\epsilon}\Big)\Big]\nonumber\\
&=&\frac{i}{2} Tr\ln \Big[\Big(1-\frac{V}{p^2(p^2-m^2)-i\epsilon
}\Big)\Big(1-\frac{V}{p^2(p^2-m^2)+i\epsilon
}\Big)-\Big(\frac{V}{p^2(p^2-m^2)}\Big)^2 \Big],
\label{317} \end{eqnarray}
where $\epsilon^2$ has been vanished. By continuing
the calculation, for this equation we obtain
\begin{eqnarray} J_{kr}&=&\frac{i}{2}Tr\ln \Big[
\Big(1-\frac{V}{p^2(p^2-m^2)-i\epsilon }\Big)
\Big(1-\frac{V}{p^2(p^2-m^2)+i\epsilon }\Big)\nonumber\\
& &\times \Big(1-\frac{(\frac{V}{p^2(p^2-m^2)})^2}{(1-\frac{V}{p^2(p^2-m^2)
 -i\epsilon})(1-\frac{V}{p^2(p^2-m^2)+i\epsilon })}\Big)\Big]\nonumber\\
&=&\frac{i}{2}Tr\ln\Big(1-\frac{V}{p^2(p^2-m^2)-i\epsilon
}\Big)+\frac{i}{2}Tr \ln\Big(1-\frac{V}{p^2(p^2-m^2)+i\epsilon
}\Big)\nonumber\\
& &+\frac{i}{2}Tr\ln\Big[1-\Big(\frac{(\frac{V}{p^2(p^2-m^2)})^2}
{(1-\frac{V}{p^2(p^2-m^2)-i\epsilon
})(1-\frac{V}{p^2(p^2-m^2)+i\epsilon })}\Big)\Big] .\end{eqnarray}  The
last term in this equation splits into two terms. So, $J_{kr}$
becomes \begin{eqnarray} J_{kr}&=&
\frac{i}{2}Tr\ln\left(1-\frac{V}{p^2(p^2-m^2)-i\epsilon
}\right)+\frac{i}{2}Tr \ln \left(1-\frac{V}{p^2(p^2-m^2)+i\epsilon
}\right)\nonumber\\ & &+\frac{i}{2}
Tr\ln\left(1-\frac{V}{p^2(p^2-m^2)-V}\right) + \frac{i}{2}
Tr\ln\left(1+\frac{V}{p^2(p^2-m^2)-V}\right).\end{eqnarray} By using the
results in appendix B, we obtain:
  \begin{eqnarray} J_{kr}&=&\frac{i}{2}Tr\ln\left(1+\frac{Y}{p^2}\right)+\frac{i}{2}Tr\ln\left(1-\frac{Y}{p^2-m^2}\right)\nonumber\\
& &+\frac{i}{2}Tr\ln\left(1+\frac{Y^2}{(m^2p^2+Y)(m^2p^2-m^4-Y)}\right)
, \end{eqnarray}
where $V=\frac{m^2 Y}{2}$ and
$ Y =2eA.P+\frac{e}{2}\sigma_{\mu\nu} F^{\mu\nu}-e^2A^2$. We
define the functions $J$, $J_0$ and $J_1$ as follow:
$$J=\frac{i}{2}Tr\ln\left(1-\frac{Y}{p^2-m^2}\right) ,\qquad J_{0}=\frac{i}{2}Tr\ln\left(1+\frac{Y}{p^2}\right)\\, $$
$$ J_{1}=\frac{i}{2}Tr\ln\left(1+\frac{Y^2}{(m^2p^2+ Y )(m^2p^2-m^4- Y )}\right),$$
so, we  have
\b J_{kr}=J+J_{0}+J_{1}.\label{eq39}\e

\setcounter{equation}{0}
\section{Regularized effective Lagrangian}

 It is clear that $J$ in (\ref{eq39}) is equivalent to the determinant fermion of Eq.(\ref{315}). In the on mass-shell scheme and for a constant electromagnetic field, the one-loop unrenormalized QED effective Lagrangian reads
\begin{eqnarray}J \equiv \delta\mathcal{L} &=&\frac{i}{2}Tr\ln\left(1-\frac{2eA.P+\frac{e}{2}\sigma_{\mu\nu}
F^{\mu\nu}-e^2A^2}{p^2-m^2+i\epsilon}\right)\nonumber\\&=&-\frac{e^2}{8\pi^2}\int_{0}^{\infty}dss^{-1}e^{-ism^2}\left[ab\coth(eas)\cot(ebs)-\frac{1}{(es)^2}
\right]\label{eq390}\end{eqnarray}
  where $a^2-b^2=\textbf{B}^2-\textbf{E}^2$, $ab=\textbf{E}\cdot\textbf{B}$ \cite{du}. The integral in (\ref{eq390}) has a divergent part proportional to $a^2-b^2$, however, to show the divergent part in the integral we use the below expansions:
   $$x\coth x\simeq 1+\frac{1}{3}x^2-\frac{1}{45}x^4+...,y\cot y\simeq 1-\frac{1}{3}y^2-\frac{1}{45}y^4+... ,$$ then
   \begin{eqnarray}xy\coth x\cot y\simeq 1+\frac{1}{3}(x^2-y^2)-\frac{1}{45}(x^2-y^2)^2-\frac{7}{45}(x^2y^2)+...,\end{eqnarray}finally, one can obtain
   \begin{eqnarray}\delta \mathcal{L} &=&-\frac{e^2}{8\pi^2}\int_{0}^{\infty}dss^{-1}e^{-ism^2}\left[-\frac{1}{3}(E^2-B^2)-\frac{e^2s^2}{45}\left((E^2-B^2)^2+7(E\cdot B)^2\right)
\right]\label{eq391},\end{eqnarray}and write
 \begin{eqnarray}\delta \mathcal{L} &=&\frac{e^2}{24\pi^2}(E^2-B^2)\int_{0}^{\infty}dss^{-1}e^{-ism^2}+\frac{2\alpha^2}{45m^4}\left[(E^2-B^2)^2+7(E\cdot B)^2\right].\label{eq392}\end{eqnarray}
 The first term while were not removed leads to a logarithmic divergence at small eigentime $s$. It's appear
 $$\int_0^{\infty}dss^{-1}\exp(-m^2s)\equiv\Gamma(0),$$
by using the following relations:
 $$\lim_{x \rightarrow
0}\Gamma(x)=\lim_{x \rightarrow 0} E_{1}(x),\qquad
E_{1}(x)=\int_{x}^{\infty}\frac{e^{-t}}{t}dt=-\gamma - \ln x - \sum_{n=1}\frac{(-1)^n x^n}{nn!},$$ where $\gamma$ is the Euler's
constant, we can write
 $$ \Gamma(0)=-\gamma-\lim_{\mu\rightarrow 0} \ln \mu ,$$
 \begin{eqnarray}\delta \mathcal{L} &=&-\frac{e^2}{24\pi^2}(E^2-B^2)(\gamma+\lim_{\mu\rightarrow 0} \ln \mu)+\frac{2\alpha^2}{45m^4}\left[(E^2-B^2)^2+7(E\cdot B)^2\right].\label{eq}\end{eqnarray}
 The divergence can be cured by $Z_3-$renormalization but we don't attempt to use the renormalization procedure. Instead, we show $J_0 $ and $J_1$ can remove the divergence at the Lagrangian. Therefore, we follow the evaluations of $J_0$ and $J_1$ in momentum representation, following the perturbation approach developed in \cite{j}, one can find

\begin{eqnarray}J_0&=&\frac{i}{2}Tr\ln\left(1+\frac{2eA.P+\frac{e}{2}\sigma_{\mu\nu}
F^{\mu\nu}-e^2A^2}{p^2+i\epsilon}\right)\nonumber\\&=&-\frac{e^2}{4\pi^2}\int_0^{\infty} dss^{-2} \int d^4k A_{\mu}(-k)A_{\mu}(k)\nonumber\\&&-\frac{e^2}{12\pi^2}\int_0^{\infty}dss^{-1}\int d^4k\frac{1}{4}F_{\mu\nu}(-k)F_{\mu\nu}(k)\nonumber\\&&+
 \frac{e^2}{16\pi^2}\int d^4kF_{\mu\nu}(-k)F_{\mu\nu}(k)\int_0^1 v^2dv \frac{1-\frac{1}{3}v^2}{1-v^2}.\end{eqnarray}
In the other hand, $J_1$ has been calculated and is expressed as below
\begin{eqnarray}J_1&\simeq & \frac{e^2}{4\pi^2}\int_0^{\infty} s^{-2} ds \int d^4k A_{\mu}(-k)A_{\mu}(k)\nonumber\\ & &+\frac{e^2}{6\pi^2}\int_0^{\infty} s^{-1}ds \int d^4k\frac{1}{4}F_{\mu\nu}(-k)F_{\mu\nu}(k)\nonumber\\ & &-\frac{e^2}{8\pi^2}\int d^4kF_{\mu\nu}(-k)F_{\mu\nu}(k) \int_0^1 dv
\frac{v^2(1-\frac{1}{3}v^2)}{1-v^2}.\end{eqnarray}
We recall that the details of $J_0$ and $J_1$  computing have been presented in Appendix A.

First, we construct  $J_{0} + J_{1}$ and have
 \begin{eqnarray}J_{01}&=& J_0+J_1 \nonumber\\
&=&\frac{e^2}{12\pi^2}\int_0^{\infty} s^{-1}ds \int d^4k\frac{1}{4}F_{\mu\nu}(-k)F_{\mu\nu}(k)\nonumber\\ & &-\frac{e^2}{16\pi^2}\int d^4kF_{\mu\nu}(-k)F_{\mu\nu}(k) \int_0^1 dv
\frac{v^2(1-\frac{1}{3}v^2)}{1-v^2}.\end{eqnarray}
This equation can be written in the following form: \b
J_{01}=\frac{e^2}{16\pi^2}\int d^4kF_{\mu\nu}(-k)F_{\mu\nu}(k)
  \left(I_{01} + I^{'}_{01}\right), \e
where

$$I_{01}=\frac{1}{3}\int_0^{\infty}s^{-1}ds,\qquad  I^{'}_{01}=-\int_0^1 dv \frac{v^2(1-\frac{1}{3}v^2)}{1-v^2}.$$
These two integrals are divergent. The divergence form of $I_{01}$ is
\begin{eqnarray} I_{01}=\frac{1}{3}\int_0^{\infty} s^{-1} ds = \frac{1}{3}\left(\lim_{\Lambda\rightarrow\infty} \ln{\Lambda}-\lim_{\mu\rightarrow 0}\ln{\mu}\right)
=-\frac{2}{3}\lim_{\mu\rightarrow 0}\ln{\mu}.\end{eqnarray}
The $I^{'}_{01}$ divergency, which has been discussed in Appendix B, is as:
\b I^{'}_{01}=-\int_0^1 dv \frac{v^2(1-\frac{1}{3}v^2)}{1-v^2}=\frac{1}{3}\lim_{\mu\rightarrow0}\ln{\mu}-\frac{1}{3}\ln2+\frac{5}{9}\label{i01}. \e
Then
\begin{eqnarray}
J_{01}=\frac{e^2}{16\pi^2}\int d^4kF_{\mu\nu}(-k)F_{\mu\nu}(k)
  \left(-\frac{1}{3}\lim_{\mu\rightarrow 0}\ln{\mu}-\frac{1}{3}\ln2+\frac{5}{9}\right).\end{eqnarray}

This expression has been obtained in the momentum representation, then one can write

\begin{eqnarray}{\mathcal{J}}_{01}=\frac{e^2}{4\pi^2}\left(-\frac{1}{4}F_{\mu\nu}F^{\mu\nu}\right)
  \left(\frac{1}{3}\lim_{\mu\rightarrow 0}\ln{\mu}+\frac{1}{3}\ln2-\frac{5}{9}\right).\end{eqnarray}
  For constant fields the Maxwell Lagrangian is given by
  $$-\frac{1}{4}F_{\mu\nu}F^{\mu\nu}=\frac{1}{2}\left(E^2-B^2\right),$$
  then we can find
  \begin{eqnarray}{\mathcal{J}}_{01}=\frac{e^2}{8\pi^2}\left(E^2-B^2\right)
  \left(\frac{1}{3}\lim_{\mu\rightarrow 0}\ln{\mu}+\frac{1}{3}\ln2-\frac{5}{9}\right).\end{eqnarray}
Therefore, $J_{kr}\equiv {\delta \mathcal{L}} _{kr}$ becomes:

\begin{eqnarray}{\delta \mathcal{L}}_{kr}&=& {\delta\mathcal{L}} +{{\mathcal{J}}_{01}} \nonumber\\ & =& \frac{e^2}{24\pi^2}(E^2-B^2)\left(-\gamma+\ln2-\frac{5}{3}\right)+\frac{2\alpha^2}{45m^4}\left[(E^2-B^2)^2+7(E\cdot B)^2\right].\label{eq393}\end{eqnarray}
Finally, we see that ${\mathcal{L}}_{kr}={\mathcal{L}}^{0}+{\delta \mathcal{L}}_{kr}$ reduces to
\begin{eqnarray} {\mathcal{L}}_{kr}&=&\frac{1}{2}(E^2-B^2)\times\left[1+
\frac{\alpha}{3\pi}\left(\ln2-\frac{5}{3}-\gamma\right)\right]\nonumber\\ & &+\frac{2\alpha^2}{45m^4}\left[(E^2-B^2)^2+7(E\cdot B)^2\right],\end{eqnarray}
 where we have added  the classical Maxwell Lagrangian, ${\mathcal{L}}^{0}=\frac{1}{2}(E^2-B^2),$ and we took $\alpha=\frac{e^2}{4\pi}$.
If one puts $\alpha=\frac{1}{137}$ and $\gamma=0.5772156649...$, the total effective Lagrangian is
\begin{eqnarray} {\mathcal{L}}_{kr}&=&(0.9987989919)\frac{(E^2-B^2)}{2}+\frac{2\alpha^2}{45m^4}\left[(E^2-B^2)^2+7(E\cdot B)^2\right],\end{eqnarray}
The result is finite and equal to the standard solution. It is easily seen that the divergent terms are disappeared and omitting each other. In fact, this is not a stochastic process but might be a natural procedure of regularization. However, to support and clarify this method and to introduce it as a successful way in regularization, we need to have some more examples of this kind.

This method of quantization may be used as an alternative way for solving the non-
renormalizability of linear quantum gravity in the background field method and is instrumental
in finding a new method of quantization, compatible with general relativity.

\section{Conclusion and outlook}

We have calculated the one-loop Euler and Heisenberg Lagrangian in the scheme of Krein space quatization with Ford's idea of fluctuated light-cone. The idea of Doubling of Hilbert space leads to the propagator expressed by the Principal Part. The other idea is the fluctuated light-cone, which is crucial for the lack of UV divergence in the results.
This is equivalent to subtracting a massless degree of freedom, and hence the UV behavior is improved similarly as Pauli-Villars.  Contrary to the Pauli-Villars case, however, the subtraction is considered physical, but the effective lagrangian will take the same form that obtained by the standard method. We remember, in the Euler-Heisenberg case, divergences occur only in the coefficients of the terms which are zeroth order and second order in field strengths, the quadratic term is nothing but the usual photon self energy term.

 As the superiority of Krein space quantization, apart from its capability on eliminating the singularity which appears in the interaction field, it should be emphasized that in presence of gauge invariance, the occurrence of Krein structure is unavoidable. Even the standard Gupta-Bleuler quantization involves a Krein space. We must emphasize the fact that this method can be used to calculate physical observable quantities
in scenarios where the effect of quantum gravity (in the linear approximation) can not be
ignored. Krein space quantization including fluctuated light-cone might be a useful candidate to achieve a quantized theory of the gravitational fields without any anomaly, as one of the greatest challenges of physics today.

\vskip 0.5 cm

\noindent {\bf{Acknowledgments}}: I would like to thank Prof. M. V. Takook  for very useful discussion and remarks.

\begin{appendix}
\setcounter{equation}{0}
 \section{Evaluating of $J_0$ and $J_1$}
 For evaluation of $J_0$ and $J_1$ we address the section 6 as "perturbation theory" in the Schwinger paper \cite{j}. The details of the evaluation based on the perturbation approach and performing in the momentum representation, will be used to find $J_0$ and $J_1$ expressions.
 In that section, after obtaining some expansions for taking the traces and doing some simplifications, the below expansion shall be retained (Eq.(6.16) in \cite{j})
 \b W^{(1)}=\frac{1}{2}ie^2\int_0^{\infty} dss^{-1}\exp(-im^2s)\times \left\{ -is Tr[A^2\exp(-ip^2s)]+ \right.$$$$ \left.
 \frac{1}{2}(-is)^2\int_{-1}^{1}\frac{1}{2}dv Tr\left[(pA+Ap) \exp\left(-ip^2\frac{1}{2}(1-v)s\right)\times (pA+Ap)\exp\left(-ip^2\frac{1}{2}(1+v)s\right)\right] \right.$$ $$ \left.
 +\frac{1}{2}(-is)^2\int_{-1}^{1}\frac{1}{2}dv Tr\left[\frac{1}{2}\sigma F  \exp \left(-ip^2\frac{1}{2}(1-v)s\right)\times \frac{1}{2}\sigma F  \exp(-ip^2\frac{1}{2}(1+v)s)\right] \right\}\label{app1}. \e
This expression for $J_0$ evaluation changes to
\b J_0=\frac{1}{2}ie^2\int_0^{\infty} dss^{-1}\exp(-im^2s)\times \left\{ +is Tr[A^2\exp(-ip^2s)]+ \right.$$$$ \left.
 \frac{1}{2}(+is)^2\int_{-1}^{1}\frac{1}{2}dv Tr\left[(pA+Ap) \exp\left(-ip^2\frac{1}{2}(1-v)s\right)\times (pA+Ap)\exp\left(-ip^2\frac{1}{2}(1+v)s\right)\right] \right.$$ $$ \left.
 +\frac{1}{2}(-is)^2\int_{-1}^{1}\frac{1}{2}dv Tr\left[\frac{1}{2}\sigma F  \exp \left(-ip^2\frac{1}{2}(1-v)s\right)\times \frac{1}{2}\sigma F  \exp(-ip^2\frac{1}{2}(1+v)s)\right] \right\}. \e
 After the end of calculations, we take $m=0$.
 For convenience, the variable $u_1$has been replaced by $\frac{1}{2}(1+v).$ The evaluation of these traces is naturally performed in a momentum representation. The matrix elements of the coordinate dependent field quantities depend only on momentum differences,
$$ \langle P+\frac{1}{2}k|A_{\mu}| P-\frac{1}{2}k\rangle=\frac{1}{(2\pi)^4}\int dx e^{-ikx}A_{\mu}(x)\equiv(2\pi)^{-2}A_{\mu}(k)$$ and
$$ \langle P|A^2_{\mu}| P\rangle=\frac{1}{(2\pi)^4}\int dxA^2_{\mu}(x)=(2\pi)^{-4}\int dk A_{\mu}(-k)A_{\mu}(k).$$
 Therefore
 \b J_0=\frac{2ie^2}{(2\pi)^4}\int_0^{\infty} dss^{-1}\exp(-im^2s)\times \left\{ +is\int d^4k A_{\mu}(-k)A_{\mu}(k)\int d^4p \exp(-ip^2s)+\right. $$$$\left.
 \frac{1}{2}(+is)^2\int_{-1}^{1}\frac{1}{2}dv \int d^4k\int d^4p 2p_{\mu}A_{\mu}(-k)\right.$$$$\left. \times \exp\left[-i\left(p+\frac{1}{2}k\right)^2\frac{1}{2}(1-v)s\right] 2p_{\nu}A_{\nu}(k)\exp\left[-i\left(p-\frac{1}{2}k\right)^2\frac{1}{2}(1+v)s\right]\right.$$$$\left.
 +\frac{1}{2}(-is)^2\int_{-1}^{1}\frac{1}{2}dv \int d^4k \int d^4p \frac{1}{4} tr\frac{1}{2}\sigma F  \right.$$$$\left. \times \exp\left(-i\left(p+\frac{1}{2}k\right)^2\frac{1}{2}(1-v)s\right) \frac{1}{2}\sigma F  \exp\left[-i\left(p-\frac{1}{2}k\right)^2\frac{1}{2}(1+v)s\right] \right\}.\e
 We thus encounter the elementary integrals
 $$ \int d^4p \exp(-ip^2s)=-i\pi^{2}s^{-2},$$
 $$ \int d^4p \exp\left[-i\left(p^2+\frac{k^2}{4}\right)s+ipkvs \right]=-i\pi^{2}s^{-2}\exp\left[-i\frac{k^2}{4}(1-v^2)s\right],$$
  $$ \int d^4p p_{\mu}p_{\nu} \exp\left[-i\left(p^2+\frac{k^2}{4}\right)s+ipkvs \right]=-i\pi^{2}s^{-2}\left(-\frac{i}{2}s^{-1}\delta_{\mu\nu}+ \frac{1}{4}v^2k_{\mu}k_{\nu}\right)\exp\left[-i\frac{k^2}{4}(1-v^2)s\right].$$

 It is convenient to replace the $\delta_{\mu\nu}$ term of the last integral by an expression which is equivalent to it in virtue of the integration with respect to $v$. Now
 $$\int_{-1}^{1}\frac{1}{2}dv\exp\left[-i\frac{k^2}{4}(1-v^2)s\right]=1-is\frac{1}{2}k^2\int_{-1}^{1}\frac{1}{3}dv v^2 \exp\left[-i\frac{k^2}{4}(1-v^2)s\right],$$
 so that, effectively
 $$ \int d^4p p_{\mu}p_{\nu} \exp\left[-i\left(p^2+\frac{k^2}{4}\right)s+ipkvs\right]=-\frac{1}{2}\pi^{2}s^{-3}\delta_{\mu\nu}$$$$+
 \frac{i}{4}\pi^2s^{-2}v^2(\delta_{\mu\nu}k^2-
 k_{\mu}k_{\nu})\exp\left[-i\frac{k^2}{4}(1-v^2)s\right].$$
 On inserting the values of the various integrals, and noticing that
 $$(\delta_{\mu\nu}k^2-k_{\mu}k_{\nu})A_{\mu}(-k)A_{\nu}(k)=\frac{1}{2}F_{\mu\nu}(-k)F_{\mu\nu}(k),$$
we obtain the immediately the gauge invariant form (with $s\rightarrow-is$)
 $$J_0=-\frac{e^2}{4\pi^2}\int_0^{\infty}ds s^{-2}\int d^4kA_{\mu}(-k)A_{\nu}(k)$$$$-\frac{e^2}{4\pi^2}\int d^4k\frac{1}{2}F_{\mu\nu}(-k)F_{\mu\nu}(k)\int_0^1 dv (1-v^2)\int_0^{\infty}ds s^{-1}\exp\left[-\left(m^2+\frac{k^2}{4}(1-v^2)\right)s\right]. $$
 This has been achieved without any special device, other than that of reserving the proper time integration to the last. A significant separation of terms is produced by a partial integration with respect to $v$, according to

 $$\int_0^1 dv (1-v^2)\int_0^{\infty}ds s^{-1}\exp[-(m^2+\frac{k^2}{4}(1-v^2))s]=\frac{2}{3}\int_0^{\infty}dss^{-1}\exp(-m^2s)-$$$$
 \frac{1}{2}k^2\int_0^1 dv (v^2-\frac{1}{3}v^4)\int_0^{\infty}ds \exp[-(m^2+\frac{k^2}{4}(1-v^2))s].$$
After calculating the last integral we obtain
\begin{eqnarray} J_0&=&-\frac{e^2}{4\pi^2}\int_0^{\infty} dss^{-2} \int d^4k A_{\mu}(-k)A_{\mu}(k)\nonumber\\&&-\frac{e^2}{12\pi^2}\int_0^{\infty}dss^{-1}\int d^4k\frac{1}{4}F_{\mu\nu}(-k)F_{\mu\nu}(k)\nonumber\\&&+
 \frac{e^2}{16\pi^2}\int d^4kF_{\mu\nu}(-k)F_{\mu\nu}(k)\int_0^1 v^2dv \frac{1-\frac{1}{3}v^2}{1-v^2}.\end{eqnarray}
Now, we want to ready the $J_1$ expansion to obtain its expression via the presented method.
In section 3 of this text, we have $J_1$ as:
\begin{eqnarray}J_1&=&\frac{i}{2}Tr\ln\left(1+\frac{Y^2}{(p^2+Y)(p^2-m^2-Y)}\right),\end{eqnarray}
we expand the logarithm function and keep the first term
\begin{eqnarray}J_1\simeq \frac{i}{2}Tr\frac{Y^2}{(p^2+Y)(p^2-m^2-Y)}
&=&-\frac{i}{2}Tr\frac{Y^2}{m^2+2Y}\left(\frac{1}{p^2+Y}-
\frac{1}{p^2-m^2-Y}\right),\end{eqnarray}
 and so we can write
\begin{eqnarray}J_1&=&-\frac{i}{2}Tr\frac{Y^2}{i(m^2+2Y)}\int_0^\infty dse^{-isp^2}\left(e^{-isY}-e^{is(Y+m^2)}\right),\nonumber\\
&=&-\frac{i}{2}Tr\frac{Y^2}{im^2}\left[1+\sum_{n=1}\left(-\frac{2Y}{m^2}\right)^n\right]\int_0^\infty
ds e^{-isp^2}\left( e^{-isY}-e^{is(Y+m^2)}\right).\end{eqnarray}
 We restrict ourselves to a specific finite number of the powers of
$Y$ to compare with the expressions of $J_{0}$, hence, we
have \b J_1\simeq \frac{i}{2}Tr\int_0^\infty s ds
Y^2e^{-isp^2}.\e
Now, in order to finding the $J_1$ expression one can use the Eq. A.1 and have:
\begin{eqnarray}J_1&\simeq & \frac{ie^2}{2}\int_0^{\infty} s ds \nonumber\\ & & \times\Big(\frac{1}{2} \int_{-1}^{1}dv Tr\Big[(p.A+A.p) \exp\Big(-ip^2\frac{1}{2}(1-v)s\Big) \times
 (p.A+A.p)\exp\Big(-ip^2\frac{1}{2}(1+v)s\Big)\Big]\nonumber\\ & &
 + \frac{1}{2}\int_{-1}^{1}dv Tr\Big[\frac{1}{2}\sigma F
 \exp\Big(-ip^2\frac{1}{2}(1-v)s\Big)\times \frac{1}{2}\sigma F  \exp\Big(-ip^2\frac{1}{2}(1+v)s\Big)\Big] \Big).\end{eqnarray}
The traces in the above expansion should be calculated in a momentum representation
 \begin{eqnarray}J_1&\simeq & \frac{2ie^2}{(2\pi)^4}\int_0^{\infty} sds \times  \Big(\frac{1}{2}\int_{-1}^{1}dv \int d^4k\int d^4p 2p.A(-k)
 \exp\Big[-i\Big(p+\frac{1}{2}k\Big)^2\frac{1}{2}(1-v)s\Big]\nonumber\\ & &\times
 2p.A(k) \exp\Big[-i\Big(p-\frac{1}{2}k\Big)^2\frac{1}{2}(1+v)s\Big]
 +\frac{1}{2}\int_{-1}^{1}dv \int d^4k \int d^4p \frac{1}{4} tr\frac{1}{2}\sigma F\nonumber\\ & &
 \times\exp\Big[-i\Big(p+\frac{1}{2}k\Big)^2\frac{1}{2}(1-v)s\Big] \frac{1}{2}\sigma F
 \exp\Big[-i\Big(p-\frac{1}{2}k\Big)^2\frac{1}{2}(1+v)s\Big]\Big),\end{eqnarray}
 then
\begin{eqnarray}J_1&\simeq & \frac{2ie^2}{(2\pi)^4}\int_0^{\infty} s ds \Big[(-2\pi^2s^{-3})\int d^4k
A_{\mu}(-k)A_{\mu}(k)\nonumber\\ & &- \int(i\pi^2s^{-2}) d^4k\frac{1}{2}F_{\mu\nu}(-k)F_{\mu\nu}(k)\int_0^1 dv(1-v^2)e^{-\frac{ik^2(1-v^2)s}{4}} \Big].\end{eqnarray}
 Finally, In the first order approximation $J_1$ expression is as
   \begin{eqnarray}J_1&\simeq & \frac{e^2}{4\pi^2}\int_0^{\infty} s^{-2} ds \int d^4k A_{\mu}(-k)A_{\mu}(k)\nonumber\\ & &+\frac{e^2}{6\pi^2}\int_0^{\infty} s^{-1}ds \int d^4k\frac{1}{4}F_{\mu\nu}(-k)F_{\mu\nu}(k)\nonumber\\ & &-\frac{e^2}{8\pi^2}\int d^4kF_{\mu\nu}(-k)F_{\mu\nu}(k) \int_0^1 dv
\frac{v^2(1-\frac{1}{3}v^2)}{1-v^2}.\end{eqnarray}

\section{Calculations}
We briefly present some calculations and simplifications which has been used in this paper.

In this section we'll bring some calculations to simplify  the logarithmic functions which was used in section(4). It is easy to see that
$$K_+=\ln \left( 1+\frac{V}{k^2(k^2-m^2)-V}\right)=-\ln\left[1-\frac{V}{k^2(k^2-m^2)}\right].$$
Then, we can write
\begin{eqnarray}\ln\left(1-\frac{V}{k^2(k^2-m^2)}\right)&=&\ln\left[1+\frac{V}{m^2}(\frac{1}{k^2}-\frac{1}{k^2-m^2})\right]\nonumber\\
&=&\ln\left[\left(1+\frac{V}{m^2k^2}\right)\left(1-\frac{V}{m^2(k^2-m^2)}\right)+\frac{V^2}{m^4k^2(k^2-m^2)}\right],\end{eqnarray}
 and finally we get to
\b  K_+=-\ln\left(1+\frac{V}{m^2k^2}\right)-\ln\left(1-\frac{V}{m^2(k^2-m^2)}\right)-\ln\left(1+\frac{V^2}{(m^2k^2+V)(m^2k^2-m^4-V)}\right).\e
Now, we calculate
 \b K_-=\ln \left( 1-\frac{V}{k^2(k^2-m^2)-V}\right)=\ln \left( \frac{k^2(k^2-m^2)-2V}{k^2(k^2-m^2)-V}\right)$$
 $$=\ln \left( 1-\frac{2V}{k^2(k^2-m^2)}\right)- \ln \left( 1-\frac{V}{k^2(k^2-m^2)}\right).\e
So, we obtain
\begin{eqnarray} K_- &=&\ln\left(1+\frac{2V}{m^2k^2}\right)+\ln\left(1-\frac{2V}{m^2(k^2-m^2)}\right)\nonumber\\ & &+\ln\left(1+\frac{4V^2}{(m^2k^2+2V)(m^2k^2-m^4-2V)}\right)
 - \ln \left( 1-\frac{V}{k^2(k^2-m^2)}\right).\end{eqnarray}

 Now, we would like to present the calculation of $I^{'}_{01}$ in (\ref{i01}):
 \b I^{'}_{01}=-\int_0^1 dv \frac{v^2(1-\frac{1}{3}v^2)}{1-v^2}= -\frac{1}{2}\int_0^1 dv\left[ \frac{v^2(1-\frac{1}{3}v^2)}{1-v}+ \frac{v^2(1-\frac{1}{3}v^2)}{1+v}\right].\e
 By using the following relations:
 $$\frac{1}{1-v}=\sum_{n=0}^{\infty}v^n,\qquad \frac{1}{1+v}=\sum_{n=0}^{\infty}(-v)^n,$$
we obtain
\begin{eqnarray}I^{'}_{01}&=&-\frac{1}{2}\int_0^1 dv \sum_{n=0}^{\infty}\left[v^{n+2}-\frac{1}{3}v^{n+4}+ (-v)^{n+2}-\frac{1}{3}(-v)^{n+4}\right]
\nonumber\\ &=&-\frac{1}{2}\sum_{n=0}^{\infty} \left[\int_0^1 dvv^{n+2}-\frac{1}{3}\int_0^1 dvv^{n+4}+ \int_0^1 dv(-v)^{n+2}-
\frac{1}{3}\int_0^1 dv(-v)^{n+4}\right]\nonumber\\
&=&-\frac{1}{2}\sum_{n=0}^{\infty} \left[\frac{v^{n+3}}{n+3}-\frac{1}{3}\frac{v^{n+5}}{n+5}+ \frac{(-1)^nv^{n+3}}{n+3}-
\frac{1}{3} \frac{(-1)^nv^{n+5}}{n+5}\right]_{0}^{1}. \end{eqnarray}
And so, we can rewrite as
$$I^{'}_{01}=-\frac{1}{2}\left\{\left[-\ln(1-v)-v-\frac{v^2}{2}\right]+\frac{1}{3}\left[\ln(1-v)+v+\frac{v^2}{2}+\frac{v^3}{3}+\frac{v^4}{4}\right]
\right.$$$$\left.
+\left[\ln(1+v)-v+\frac{v^2}{2}\right]-\frac{1}{3}\left[\ln(1+v)-v+\frac{v^2}{2}-\frac{v^3}{3}+\frac{v^4}{4}\right]\right\}_{0}^{1} .$$
Finally, we have
\begin{eqnarray} I^{'}_{01}&=&-\left[-\frac{1}{3}\ln(1-v)+\frac{1}{3}\ln(1+v)-\frac{2}{3}v+\frac{1}{9}v^3\right]_{0}^{1}
\nonumber\\ &=&\frac{1}{3}\lim_{\mu\rightarrow0}\ln{\mu}-\frac{1}{3}\ln2+\frac{5}{9}.
 \end{eqnarray}

\end{appendix}


\begin{thebibliography}{a}

\bibitem{itzu} C. Itzykson, J. B. Zuber, McGraw-Hill, Inc. (1988) {\it Quantum Field
Theory}.
\bibitem{wehais}S. Weinberg, {\it Ultraviolet divergences in quantum gravity theories, in:
General relativity}, Eds. S. W. Hawking and W. Israel, cambridge Univ. Press, (1979)
\bibitem{iga}I. G. Avramidi, Covariant Methods for the Calculation of
the Effective Action in Quantum Field Theory and Investigation of Higher-Derivative
Quantum Gravity, hep-th/9510140v3.
\bibitem{gazeau00} J. P. Gazeau, J. Renaud, M. V. Takook, Class. Quantum
Grav. $17, 1415(2000)$, gr-qc/$9904023$.
\bibitem{ta3} M. V. Takook, Mod. Phys. Lett. A, 16(2001)1691,
gr-qc/0005020.
\bibitem{ta4} M. V. Takook, Int. J. Mod. Phys. E, 11(2002)509,
gr-qc/0006019.
\bibitem{rota} S. Rouhani, M. V. Takook, Int. J. Theor. Phys., 48(2009)2740–2747.
\bibitem{gagarota} T. Garidi et al, J. Math. Phys., 49(2008)032501.
\bibitem{gaj} T. Garidi et al, J. Math. Phys., 44(2003)3838.
\bibitem{sb} S. Behroozi et al, Phys. Rev. D, 74(2006)124014.
\bibitem{derotata} M. Dehghani et al, Phys. Rev. D, 77(2008)064028.
\bibitem {taj}M. V. Takook et al, J. Math Phys., 51(2010)032503.
\bibitem{reta} A. Refaei, M. V. Takook, Mod. Phys. Lett. A, 26(2011)31, arXiv: 1109.2693.
\bibitem{reta2} A. Refaei, M. V. Takook, Phys. Lett. B, 704(2011)326, arXiv: 1109.2692.
\bibitem{des} S. Deser, Rev. Mod. Phys., 29(1957)417.
\bibitem{dew} B. S. DeWitt, Phys. Rev. Lett., 13(1964)114.
\bibitem{ford97} H. L. Ford, Quantum Field Theory in Curved Spacetime,
gr-qc/9707062.
\bibitem{dir}P. A. M. Dirac, Proc. Roy. Soc. A, 180(1942)1
\bibitem{rami}  A. Ramirez, Mielnik B., Rev. Mex. Fis., 49S2(2003) 130, quant-ph/0211048.
\bibitem{gu}  S. N. Gupta, Proc. Phys. Soc. Sect. A, 63(1950)681
\bibitem{haw} S. W. Hawking, T. Hertog, Phys. Rev. D, 65(2002)103515
\bibitem{ha} A. D. Helfer, The Physics of Negative Energy Densities, hep-th/9811081
\bibitem{viba}M. Visser, C. Barcelo, Energy condition and their cosmological implications, gr-qc/0001099
\bibitem{bl} S. Blinnikov, Surveys High Energ. Phys. 15(2000)37, astro-ph/9911138
\bibitem{allen85} B. Allen, Phys. Rev. D, $32 (1985) 3136$.
\bibitem{heeu} W. Heisenberg and H. Euler, Consequences of Dirac's theory of positrons," Z. Phys. 98, 714, (1936).
\bibitem{j} J. Schwinger, Phys. Rev. 82(1951)664.
\bibitem{geduad} V. Dunne, Adolfo Huet, Jin Hur, Hyunsoo Min, The derivative expansion at small mass for the spinor effective action, hep-th/1103.3150.
\bibitem{du} G. V. Dunne, Heisenberg-Euler effective Lagrangians: Basics and extensions," in Ian Kogan Memorial Collection, From
Fields to Strings: Circumnavigating Theoretical Physics', M. Shifman et al, World Scientific Publ., Singapore, (2005), hep-th/0406216.
\bibitem{bida} N. D. Birrell, P.C.W. Davies, Cambridge University Press (1982) {\it Quantum Fields in Curved Space}.
\bibitem{bach}  N. H. Barth, S.M. Christensen,
 Phys. Rev. D, 28(1983)1876.
 \bibitem{ho} P. Horava, Phys. Rev. D, 79(2009)084008, arXiv:0901.3775.
\bibitem{ka} M. Kaku, Oxford University Press, (1993) {\it Quantum Field
Theory}.


\end{thebibliography}
\end{document}